\documentstyle{article}
\begin{document}

\title{
Effective Potential at Finite Temperature\footnote{
Talk given at YITP International Workshop on Physics of
Relativistic Heavy Ion Collisions, Kyoto, Japan, 9--11
June 1997, to be published in Progress of Theoretical
Physics Supplement (Proceedings).} 
 \\
{\small
--- RG improvement vs. High temperature expansion ---
}
}
\author{
Hisao Nakkagawa\thanks{E-mail:~nakk@daibutsu.nara-u.ac.jp} 
and Hiroshi
Yokota\thanks{E-mail:~yokotah@daibutsu.nara-u.ac.jp} \\
{\small\it Institute for Natural Science, Nara University, 1500
Misasagi-cho,} \\
{\small\it Nara 631, Japan}
}
\date{}

\maketitle

\begin{abstract}
We have applied the recently proposed renormalization group
improvement procedure of the finite temperature effective potential,
and have investigated extensively the phase structure of the massive
scalar $\phi^4$ model, showing that the $\phi^4$ model has a rich 3-phase
structure at $T \neq 0$, two of them are not
seen in the ordinary perturbative analysis.  Temperature dependent phase
transition in this model is shown to be strongly the first order.
\end{abstract}

\vspace{0.5cm}
\hspace{-0.8cm}
{\it 1. Introduction}:\ \ 
Recently in the analysis of the effective potential (EP) 
simple but efficient procedures for resumming dominant large
corrections appeared in the perturbative calculation
by using the renormalization group (RG) technique thus
resolving also the problem of renormalization scheme (RS)
dependence\cite{rf:2}
are proposed in vacuum\cite{rf:3} and in thermal\cite{rf:4}
field theories.  In this paper we briefly present the result of application
of this procedure to the massive scalar $\phi^4$ model at finite temperature,
showing that this model has three phases, only
one of them can be seen in the ordinary perturbative analysis. Other two
phases emerge as a result of resummation of dominant large corrections, one
of them has a truly non-perturbative nature and can not be seen in the high
temperature expansion analysis.
Temperature dependent phase transition between the ordinary and the new
phases proceeds through 
the strong first order transition.  Details of our analyses will be given
elsewhere.\cite{rf:5} 

\hspace{-0.8cm}
{\it 2. RG improvement procedure}:\ \ 
Let us focus on the massive self-coupled scalar $\phi^4$ model at finite
temperature,
\begin{equation}
  {\cal L} = (\partial_{\mu} \phi )^2/2 - m^2 \phi^2/2
      - \lambda \phi^4/4! - h m^4, \ \ (m^2 < 0).
\end{equation}
The exact EP satisfies
a renormalization group equation (RGE),
\begin{equation}
     \left( \mu \frac{\partial}{\partial \mu} +
        \beta \frac{\partial}{\partial \lambda} -
        m^2 \theta \frac{\partial}{\partial m^2} -
        \phi \gamma \frac{\partial}{\partial \phi} +
        \beta_{h} \frac{\partial}{\partial h} \right) V =0,
\end{equation}
whose solution is
\begin{equation}
     V \left( \phi, \lambda, m^2, h, T; \mu^2 \right) =
     \bar{V} \left( \bar{\phi}(t), \bar{\lambda}(t), \bar{m}^2(t),
        \bar{h}(t), T; \bar{\mu}^2=\mu^2 e^{2t} \right),
\end{equation}              
where the barred quantities $\bar{\phi}$, $\bar{\lambda}$ etc. are running
parameters
whose responces to the change of $t$ are determined by the coefficient
functions of the RGE, $\gamma$, $\beta$ etc., with the boundary
condition that the barred quantities reduce to the unbarred parameters at
$t = 0$. The problem of resolving the RS-dependence
of the EP now reduces the one how we can determine, with the limited knowledge
of the $L$-loop calculation, the function form of the EP at least at a certain
value of $t$.

Let us notice here that in the scalar $\phi^4$ model (at least in the
$O(N)$ symmetric model in $N \to \infty$) the dominant large corrections
appear as power functions of the effective variable 
$\tau \equiv \lambda \Delta_1$, 
having the high temperature behavior
$\tau \sim \lambda T^2/M^2$, where $M^2=m^2+\lambda \phi^2/2$. We can see
\cite{rf:4} that the EP can be expressed in the power-series expansion in
$\tau$;
\begin{equation}
    V =  \frac{M^4}{\lambda}
     \sum_{\ell =0}^{\infty} \lambda^{\ell} \: \left[ \: 
               F_{\ell}(\tau) + z \delta_{\ell ,0}
            \:  \right] \ , \ \ \ z \equiv \frac{\lambda h m^4}{M^4} ,
\end{equation}
where
\begin{equation}
     F_{\ell}(\tau) \equiv  \sum_{L=\ell}^{\infty} v_{\ell}^{(L)}
             \tau^{L- \ell} \ , \ \ F_{0}(\tau) = \sum_{L=0}^{2} v_{0}^{(L)} \tau^{L}\ , 
\ \ v_0^{(L)}=0 \mbox{ for } L \ge 3 \ .
\end{equation}
By remembering the fact that,
at $\tau =0$, the ``$\ell$ th-to-leading $\tau$'' function $F_{\ell}$ is given
solely by the $\ell$-loop level potential,
$F_{\ell}(\tau=0)=v_{\ell}(L=\ell)$, we find
if we caluculated the EP to the $L$-loop level, then at $\tau = 0$ it already
gives the function form ``exact'' up to ``$L$th-to-leading $\tau$'' order.
With the $L$-loop potential at hand, the EP satisfying the RGE can then
be given by
\begin{eqnarray}
     V &=& \bar{M}^4(t)  \left.
      \sum_{\ell =0}^{L} \bar{\lambda}^{\ell -1}(t)
       \left[ \: \bar{v}_{\ell}^{(\ell)}(t) +
            \bar{z}(t) \delta_{\ell,0} \:
            \right] \right|_{\bar{\tau}(t)=0}     \nonumber  \\
   &=&  V_L  \left. ( \phi, \bar{\lambda}(t),
             \bar{m}^2(t), \bar{h}(t);
            \mu^2 e^{2t}) \right|_{\bar{\tau}(t)=0} \ ,
\end{eqnarray}
where the barred quantities should be evaluated at such a $t$ satisfying
$\bar{\tau}(t)=0$.

\hspace{-0.8cm}
{\it 3. Phase structure of the massive scalar $\phi^4$ model at $T \neq 0$}:\ \ 
Now we apply the RG improvement procedure explained
above to the massive scalar $\phi^4$ model at $T \neq 0$, and study the phase
structure. The perturbatively calculated one-loop EP is
\begin{eqnarray}
 V_1 = & & m^2 \phi^2/2 + \lambda\phi^4/4! + hm^4 \nonumber \\
       & + &  \frac{M^4}{2} \left[ \tau 
          + \lambda \left\{ - \frac{1}{64\pi^2} +
         \frac{T^4}{\pi^2 M^4} L_0 \left( \frac{T^2}{M^2} \right)
        - \frac{T^2}{2\pi^2M^2}
        L_1 \left(\frac{T^2}{M^2} \right) \right\} \right] , 
\end{eqnarray}
where
\begin{eqnarray}
    \tau & \equiv &
            \lambda \left\{ \frac{1}{32\pi^2} \left(
        \ln \frac{M^2}{\mu^2} -1 \right) + \frac{T^2}{2\pi^2M^2}
        L_1 \left(\frac{T^2}{M^2} \right) \right\}  \ , \\
    L_0 \left( \frac{1}{a^2} \right) &\equiv &
       \int_0^{\infty} k^2 \, dk \, \ln [ 1 - \exp \{ - \sqrt{k^2+a^2} \} ]
       \ , \ \ 
    L_1 \left( \frac{1}{a^2} \right) \equiv  \frac{\partial}{\partial a^2}
       L_0 \left( \frac{1}{a^2} \right) . \ \ \ 
\end{eqnarray}
At the one-loop level the RGE coefficient functions are
$\gamma =1$, $\beta = b \lambda^2$, $\theta = -b \lambda$, $\beta_h = b/2 -
2bh\lambda$, where $b=1/16\pi^2$. Thus the RG improvement can be performed
analytically, obtaining the improved EP by Eq.(7) with all the renormalized
parameters replaced by the barred quantities.
High temperature expansion then gives
\begin{equation}
  \bar{V}_1 (t) 
      = \frac{1}{2} \bar{m}^2 \phi^2 + \frac{1}{4!} \bar{\lambda}\phi^4
         - \frac{\bar{m}^4}{2 \bar{\lambda}} + 
          \frac{T^2 \bar{M}^2}{48} - \frac{T \bar{M}^3}{48 \pi} 
         + \cdots  ,
\end{equation}   
where
\begin{equation}
   \bar{M}^2(t) =  \bar{m}^2 + \bar{\lambda} \phi^2/2, \ \ 
   \bar{\lambda}(t) =  \lambda (1 - 3 \lambda b t)^{-1} , \ \
   \bar{m}^2(t) = m^2 (1 - 3 \lambda b t)^{-1/3} 
\end{equation}
and all the barred quantities should be evaluated at such a $t$ satisfying
the RS-fixing condition $\bar{\tau}(t)=0$, which gives the mass gap
equation,\cite{rf:6}
\begin{equation}
  M^2= m^2 + \bar{M}^2 f(\bar{M}^2) -  m^2 f(\bar{M}^2)^{2/3}, 
\end{equation}
\begin{equation}
  f(\bar{M}^2) = 1 - 3 \lambda b t = 1 - 3 \lambda \left[
     \frac{T^2}{24 \bar{M}^2} - \frac{T}{8 \pi \bar{M}}
     + b \{ \ln (\frac{4 \pi T}{\mu}) - \gamma_E \} + \cdots \right].
\end{equation}

The stationary condition $d \bar{V}_1/d \phi=0$ gives
$ \phi ( \bar{M}^2 - \bar{\lambda} \phi^2/3 )= 0$.
With the above RG improved formula in hand we can see the
phase structure of the model.
 
\hspace{-0.8cm}
{\it 3.1 High temperature expansion analysis}:\ \ 
First let us see the result in the high temperature expansion. In
this case the mass gap equation gives at sufficiently high
temperature the $\phi^2$-$\bar{M}^2$ relation
shown in Fig.1, indicating
the existence of two phases I and II. This is
shown to be what really happened, see Fig.2, in which we can see that
the phase I is the symmetric phase and the phase II is the broken one.
Also we can see is that at low temperature below $T_1$ the 
broken phase realizes the true vacuum,
but that as the temperature becomes
higher the symmetric and the broken phases eventually become mixed up thus
showing the bump structure in the potential and finally at high temperature
above $T_3$ the symmetric phase I realizes the true vacuum.
Phase transition in
this case is strongly first order. The broken phase II is the ordinary one 
being related to the tree EP, while the symmetric phase I is generated by the
resummation effect of the large-$T$ ($T^2$) corrections.
It is worth noting that in both phases I and II the running parameters
$\bar{\lambda}$ and $\bar{m}^2$ may show some peculiar behaviors, i.e., in the 
small
$\phi$ region they blow up, and in the phase I $\bar{\lambda}$ becomes negative.
These are not the real trouble because if we correctly define the effective
coupling and the effective mass-squared by
$\lambda_{eff} = d^4 \bar{V}_1/d \phi^4$ and 
$m^2_{eff} = d^2 \bar{V}_1/d \phi^2$,
then $\lambda_{eff}$ and $m^2_{eff}$ show moderate behavior being consistent
with the perturbative treatment, except in the very small $\phi$ region where
$\lambda_{eff}$ becomes negative. This result may be related with the small
$\phi$ problem pointed out by Amelino-Camelia.\cite{rf:7} \ 
More detailed study on this problem will be given in Ref.~4).

\hspace{-0.8cm}
{\it 3.2 Full analysis with RG improved $\bar{V}_1$ and $\tau$}:\ \ 
The ``exact'' mass gap equation $\bar{\tau}(t)=0$ gives the
$\phi^2$-$\bar{M}^2$ relation shown in Fig.3, indicating the existence of
three phases: two of them, i.e.,
phases I and II are those already appeared in the high temperature
expansion analysis, whereas the third phase III is totally new.
In this phase the effective coupling $\lambda_{eff}$ becomes strong and the 
effective mass-squared also becomes very heavy, indicating this phase to be
almost temperature independent super massive strong coupling phase.

\hspace{-0.8cm}
{\it 4. Discussion and comments}:\ \ 
In this paper we present briefly the RG improved analysis of the
effective potential in massive scalar $\phi^4$ model at $T\neq0$.
The same analysis can be done in the same model at $T= 0$,
disclosing\cite{rf:5} that the $T \to 0$ limit
of the model at $T\neq0$ does not 
coincide with the same model at $T= 0$. The $T\to0$ limit of the model
at $T\neq0$ maintains the same ``stable'' phase structure
as those at $T\neq0$, while the model at $T= 0$ shows the unstable phase
structure, i.e., the couterpart of the phase I becomes unbounded from below
thus the theory becomes unstable.

One comment on the small $\phi$ region problem is added.
Amelino-Camelia\cite{rf:7}
noted that the small $\phi$ region reliability may be related with the types of
resummed diagramms.
Because of the simplicity of the model we can explicitly count the types of
diagramms being resummed through the improvement procedure, and find that
the resummed graphs are chains, daisies, and the dominant superdaisies,
suggesting the reliability range to be $\phi$ roughly greater equal to
$\sqrt{\lambda} T$. This range may be consistent with the region of $\phi$ where
$\lambda_{eff}$ becomes positive and moderate, see
Ref.~4).

\vspace{1cm}
\centerline{\bf Figure Caption}
\begin{description}
\item[Fig.1:]
$\phi^2$-$\bar{M}^2$ relation from the High temperature expression
of the mass gap equation, $\bar{\tau}(t)=0$, Eqs.(11)--(13).
\item[Fig.2:]
RG improved effective potentials at three temperatures:
(a) $\tilde{T}_1= 20.0$, (b) $\tilde{T}_2= 23.9$, and
(c) $\tilde{T}_3= 25.0$. $\tilde{V} \equiv \bar{V}_1(\phi) - 
min\{\bar{V}_1(\phi=0)\}$, $\tilde{T} \equiv T/|m|$, and the coupling is set
to $\lambda = 1/20$.
\item[Fig.3:]
$\phi^2$-$\bar{M}^2$ relation from the ``exact'' mass gap equation,
$\bar{\tau}(t)=0$, Eqs.(11)--(12).
\end{description}

\begin{thebibliography}{99}
\bibitem{rf:2}
See, e.g., T.~Muta, {\it Foundations of Quantum Chromodynamics}
(World Scientific, 1987).
\bibitem{rf:3}
M.~Bando, T.~Kugo, N.~Maekawa and H.~Nakano, 
        Phys.~Lett.\ {\bf B301} (1993), 83.
\bibitem{rf:4}
H.~Nakkagawa and H.~Yokota, Mod.~Phys.~Lett. {\bf A11} (1996), 2259.
\bibitem{rf:5}
H.~Nakkagawa and H.~Yokota, three papers to appear (1997).
\bibitem{rf:6}
G.~Amelino-Camelia and S.-Y.~Pi, Phys. Rev. {\bf D47} (1993) ,2356.
\bibitem{rf:7}
G.~Amelino-Camelia, Phys. REv. {\bf D49} (1994), 2740.
\end{thebibliography}
\end{document}